\documentclass[aps, prd, amsmath, floats, floatfix, twocolumn, nofootinbib, superscriptaddress]{revtex4-1}
\usepackage{graphicx}
\usepackage{color}

\newcommand{\be}{\begin{eqnarray}}
\newcommand{\ee}{\end{eqnarray}}
\newcommand{\rar}{\rightarrow}

\begin{document}

\title{Scattering of particles by deformed non-rotating black holes}

\author{Guancheng Pei}
\affiliation{Center for Field Theory and Particle Physics and Department of Physics, Fudan University, 200433 Shanghai, China}

\author{Cosimo Bambi}
\email[Corresponding author: ]{bambi@fudan.edu.cn}
\affiliation{Center for Field Theory and Particle Physics and Department of Physics, Fudan University, 200433 Shanghai, China}
\affiliation{Theoretical Astrophysics, Eberhard-Karls Universit\"at T\"ubingen, 72076 T\"ubingen, Germany}

\date{\today}

\begin{abstract}
We study the excitation of axial quasi-normal modes of deformed non-rotating black holes by test-particles and we compare the associated gravitational wave signal with that expected in general relativity from a Schwarzschild black hole. Deviations from standard predictions are quantified by an effective deformation parameter, which takes into account deviations from both the Schwarzschild metric and the Einstein equations. We show that, at least in the case of non-rotating black holes, it is possible to test the metric around the compact object, in the sense that the measurement of the gravitational wave spectrum can constrain possible deviations from the Schwarzschild solution.
\end{abstract}

\maketitle

%%%%%%%%%%%%%%%%%%%%%%%%%%%%%%%

\section{Introduction}

So far, general relativity has successfully passed all the experimental tests~\cite{will}. However, its most interesting predictions are still to be verified. The theory has been tested in weak gravitational fields, mainly with precise experiments in the Solar System and accurate radio observations of binary pulsars. The agreement between theoretical predictions and observational data is today confirmed with a precision ranging from a few percent to about $10^{-5}$. There is now an increasing interest to test the theory in other regimes, in particular at very large scales and in strong gravitational fields. Tests of general relativity at very large scales are relevant for the problems of dark energy and dark matter: in the last 20 years, this has been a very active research field~\cite{cosmo}. The ideal laboratory to test strong gravity is the spacetime around astrophysical black hole (BH) candidates~\cite{bh1,bh2}. The spacetime geometry around these objects has still to be verified and there is a number of theoretical arguments suggesting that macroscopic deviations from standard predictions are possible~\cite{new1,new2,new3}.

In 4-dimensional general relativity, uncharged BHs are described by the Kerr solution and they are completely specified by only two parameters, namely the mass $M$ and the spin angular momentum $J$. This is the well-known Òno-hairÓ theorem~\cite{hair1,hair2}. Astrophysical BHs should form from the collapse of very massive stars, after the latter have exhausted all their nuclear fuel~\cite{luca}. According to the theory of general relativity, the final product of the collapse should be well described by the Kerr solution. Initial deviations from the Kerr metric are quickly radiated away through the emission of gravitational waves~\cite{collapse}. For macroscopic objects, the equilibrium electric charge is extremely small and is reached soon because of the highly ionized host environment~\cite{charge}. The effect of an accretion disk is normally negligible, as the disk mass is many orders of magnitude smaller than the mass of the central object~\cite{disk}.

Compact objects in X-ray binaries are classified as stellar-mass BH candidates if they exceed 3~$M_\odot$, which is the maximum mass expected for neutron and quark stars~\cite{rr}. The supermassive objects at the center of every normal galaxy are called supermassive BH candidates because they cannot be explained as clusters of non-luminous objects: the expected cluster lifetime due to evaporation and physical collisions would be shorter than the age of these systems~\cite{maoz}. The non-observation of thermal radiation emitted by the possible surface of these objects may also be interpreted as an evidence for an event/apparent horizon~\cite{eh1} (but see Refs.~\cite{eh2,eh3} and there are also scenarios in which no thermal component should be expected~\cite{new1,new2}). In the end, both stellar-mass and supermassive BH candidates can be naturally interpreted as the Kerr BHs of general relativity and they could be something else only in presence of new physics. However, a direct observation confirmation is completely missing.

The nature of BH candidates can be potentially tested by studying the properties of the radiation emitted by the accreting gas~\cite{cb1,cb2,cb3,cb4,cb5,b055,jp1,jp2}. The electromagnetic spectrum of these objects has indeed features determined by the motion of the gas in the accretion disk and by the propagation of the photons from the disk to the distant observer. The study of these features can potentially constrain the spacetime geometry and check whether these objects are the Kerr BHs of general relativity. Current observations cannot do it, mainly because there is a fundamental degeneracy among the parameters of these systems: the relativistic features produced in a non-Kerr background cannot be distinguished from those produced around a Kerr BH with a different spin~\cite{corr1,corr2,corr3,corr4}. With the available data, we can rule out some exotic BH alternatives, like some kinds of compact objects~\cite{e1,e2} and of traversable wormholes~\cite{e3}. Non-Kerr objects with a horizon are much more difficult to test. Current constraints on stellar-mass BH candidates from the disk's thermal spectrum are reported in~\cite{cfm1,cfm2}. The possibility of testing the Kerr metric in the future seems to depend on the possibility of combining several observations of the same source to break the degeneracy among the parameters of the system~\cite{f1,f2}.

A complementary approach, not yet available, is to use the gravitational wave signal~\cite{gw1,gw2,gw3}. This approach has advantages and disadvantages. Assuming geodesic motion, the properties of the radiation emitted by the gas in the accretion disk only depend on the background metric, and therefore they can be used to test the Kerr metric. They cannot distinguish a Kerr BH in general relativity from a Kerr BH in an alternative theory of gravity~\cite{psaltis}. On the contrary, the emission of gravitational waves depends on both the background metric and the field equations of the theory, with the advantage that it may be possible to distinguish Kerr BHs in different theories and the disadvantages that a more rigorous treatment would require the knowledge of the field equations. However, a phenomenological description of the problem is possible and it has been already employed, for instance, in Refs.~\cite{gwp1,gwp2}.

In this work, we investigate the gravitational wave signal emitted when the axial quasi-normal modes of a BH are excited by the passage of a test-particle. As an exploratory work, we consider the simplest case of non-rotating BHs. We derive the master equations and we compute the wave form and the energy spectrum as a function of the deviations from general relativity and of the energy and the angular momentum of the test-particle. Deviations from standard predictions are quantified by an effective deformation parameter, which is used to take into account a possible non-Schwarzschild background and corrections to the Einstein equations. We find that the gravitational wave signal is mainly determined by the geodesic motion of the test-particle, while the contribution from the proper excitations of the spacetime is smaller.

The content of the paper is as follows. In Section~\ref{s-sch}, we review the excitation of axial quasi-normal modes of a Schwarzschild BH by a test-particle in general relativity. In Section~\ref{s-jp}, we consider the same phenomenon in the case of a non-rotating deformed BH. Since we compute the gravitational wave signal from the Einstein equations, eventually our predictions are parametrized by an ``effective'' deformation parameter. Such a deformation parameter is not the same as the one appearing in the background metric and therefore it cannot be directly compared with the deformation parameter constrained by the studies of the electromagnetic spectrum. In Section~\ref{s-sim}, we present the results of our calculation and we compare the gravitational wave signals from Schwarzschild and deformed non-rotating BHs. Summary and conclusions are reported in Section~\ref{s-c}. Throughout the paper, we use units in which $G_{\rm N} = c = 1$.

\vspace{0.5cm}

\section{Scattering of particles by a Schwarzschild black hole \label{s-sch}}

To study the gravitational wave signal by a test-particle perturbing the Schwarzschild background, we can proceed as follows~\cite{kk1}. At the zeroth order, the Einstein equations are $R^0_{\alpha\beta}=0$ and the unperturbed solution is the static Schwarzschild background $g^0_{\alpha\beta}$. The first order term in the perturbed metric is $g^1_{\alpha\beta}$ and it can be derived by
\be\label{eq-ee1}
R^1_{\alpha\beta} = 8\pi \left(T_{\alpha\beta} - 
\frac{1}{2} g^0_{\alpha\beta} T^{\gamma}_{\gamma} \right) \, .
\ee
We expand the metric $g^1_{\alpha\beta}$ in tensor spherical harmonics. In this work, we only consider axial perturbations. The equations governing polar perturbations can be obtained from the axial ones by a coordinate transformation. We choose the so-called Regge-Wheeler gauge and the perturbed line element reads~\cite{kk1}
\begin{widetext}
\be
ds^2 = g^0_{\alpha\beta}dx^\alpha dx^\beta
&+& 2\sin\theta \sum_{l,m} \frac{\partial Y_{lm}(\theta,\phi)}{\partial\theta}
\left[ f_{lm}(t,r) dt d\phi + h_{lm}(t,r) dr d\phi\right] \nonumber\\
&-& \frac{2}{\sin\theta} \sum_{l,m} 
\frac{\partial Y_{lm}(\theta,\phi)}{\partial\phi}
\left[ f_{lm}(t,r) dt d\theta + h_{lm}(t,r) dr d\theta\right] \, .
\ee
The matter source in the Einstein equations is given by the stress-energy tensor of a point-like particle moving along the geodesics of the Schwarzschild background, namely 
\be
T^{\alpha\beta} = \mu \frac{dT}{d\tau} \frac{dz^\alpha}{dt} 
\frac{dz^\beta}{dt} \frac{1}{r^2} \delta\left(r - R(t)\right) 
\delta\left(\theta - \Theta(t)\right) \delta\left(\phi - \Phi(t)\right) \, ,
\ee
where $\mu$ is the particle's rest-mass, $z^\alpha = \left(T(t), R(t), \Theta(t), \Phi(t)\right)$ is the world-line trajectory of the particle in term of the Schwarzschild $t$-coordinate, and $\tau$ is the particle's proper time. The tensor $T^{\alpha\beta}$ is expanded in tensor spherical harmonics as well~\cite{zerilli}. After some tedious calculations, the components $t\phi$, $r\phi$, and $\theta\phi$ of Eq.~(\ref{eq-ee1}) provide the following equations
\be\label{eq-axial}
&&\frac{\partial^2}{\partial t^2} Z_{lm}(t,r) - \frac{\partial^2}{\partial r_*^2} Z_{lm}(t,r)
+ V_l(r) Z_{lm}(t,r) = S_{lm}(t,r) \, , \\
&&\frac{\partial}{\partial t} f_{lm}(t,r) = 
\frac{\partial}{\partial r_*} \left[r_* Z_{lm}(t,r) \right] \, ,
\ee
where
\be
Z_{lm}(t,r) = \left(1 - \frac{2M}{r}\right)\frac{h_{lm}(t,r)}{r}
\ee
and $r_*$ is the tortoise coordinate, namely
\be
r_* = r + 2M \ln \left(\frac{r}{2M} - 1\right) \, .
\ee 
The potential $V_l(r)$ and the source term $S_{lm}(t,r)$ are given, respectively, by
\be
V_l(r) &=& \left(1 - \frac{2M}{r}\right)
\left[\frac{l(l+1)}{r^2} - \frac{6M}{r^3}\right] \, , \\
S_{lm}(t,r) &=& - 16\pi i \left(1 - \frac{2M}{r}\right)
\left\{r \frac{\partial}{\partial r}\left[\left(1 - \frac{2M}{r}\right) D_{lm}\right] - 
\left(1 - \frac{2M}{r}\right) Q_{lm}\right\} \, ,
\ee
where
\be
D_{lm} &=& - \frac{\mu m}{l(l+1)(l-1)(l+2)} \frac{L_z^2}{E} \frac{1}{r^4}
\left(1 - \frac{2M}{r}\right) \delta\left(r - R(\tau)\right) 
\frac{\partial Y^*_{lm}}{\partial\theta} \Big|_{\Theta=\frac{\pi}{2} , \Phi(t)} \, , \\
Q_{lm} &=& - \frac{i \mu}{l(l+1)} \frac{L_z}{r^2(r - 2M)} \frac{dR}{d\tau} 
\delta\left(r - R(\tau)\right) \frac{\partial Y^*_{lm}}{\partial\theta}
\Big|_{\Theta=\frac{\pi}{2} , \Phi(t)}\, .
\ee
Here $E$ and $L_z$ are, respectively, the specific energy and the axial component of the specific angular momentum of the particle, which is assumed to move on the equatorial plane $\Theta = \frac{\pi}{2}$. The wave form of the signal is $Z_{lm}(t) = \lim_{r\rar+\infty} Z_{lm}(t,r)$. The energy spectrum of gravitational waves at infinity is proportional to
\be\label{eq-dedo}
\frac{dW}{d\omega} = \frac{1}{16\pi^2} \sum_{l=2}^{+\infty}
\sum_{m=-l}^{l} (l-1)l(l+1)(l+2) \left| Z_{lm}(\omega) \right|^2 \, ,
\ee
\end{widetext}
where $Z_{lm}(\omega)$ is the Fourier transform of the wave form $Z_{lm}(t)$. We remind the reader that there are no monopole $(l=0)$ and dipole $(l=1)$ contributions in general relativity, and that the leading order term is the quadrupole $(l=2)$ one. Moreover, multipoles with $m=l-1$, $l-3$, etc. correspond to axial perturbations, while those with $m=l$, $l-2$, etc. correspond to polar perturbations. As a preliminary study, here we are restricting our attention to axial perturbations only, and therefore we will ignore the signal associated to polar perturbations.

\section{Scattering of particles by a deformed black hole \label{s-jp}}

In the case of an alternative theory of gravity, one could proceed in the same way, namely starting from the corresponding BH solution and considering the perturbations generated by a test-particle moving along the geodesics of that metric (in the case of a metric theory of gravity). The evolution of these perturbations and the associated gravitational wave signal at infinity are then determined by the field equations of the theory. The final result depends on both the background metric and the field equations. In this sense, the approach is more powerful than the study of the properties of the radiation emitted by the gas in the accretion disk, as the latter is only sensitive to the background metric. However, if we use gravitational waves, we can only test a particular theory, while the actual fundamental theory may not be known.

In this paper, we explore a different approach. We consider the Johannsen-Psaltis metric~\cite{jpm}, which describes the gravitational field around non-Kerr BHs and in which the deviations from the Kerr geometry are quantified by a set of ``deformation parameters''. The simplest non-rotating BH has only one non-vanishing deformation parameter and its line element reads
\be\label{gmn}
\hspace{-1cm}
ds^2 &=& - \left(1 - \frac{2 M}{r}\right) 
\left(1 + \epsilon \frac{M^3}{r^3}\right) dt^2 \nonumber\\
&& + \left(1 - \frac{2 M}{r}\right)^{-1} \left(1 + \epsilon \frac{M^3}{r^3} \right) dr^2
+ r^2 d\Omega \, ,
\ee
where $\epsilon$ is the deformation parameter and the gravitational force is weaker (stronger) than the one around a Schwarzschild BH with the same mass when $\epsilon > 0$ ($\epsilon < 0$). If $\epsilon = 0$, we recover the Schwarzschild solution. We note that the horizon is at $r_H = 2M$ for any value of $\epsilon$. The metric in Eq.~(\ref{gmn}) is not a solution of any known field equations of an alternative theory of gravity. It is just the Schwarzschild metric with an ad hoc deformation, which is quantified by the parameter $\epsilon$. Like the parameters $\beta$ and $\gamma$ in the Parametrized Post-Newtonian formalism extensively employed to test general relativity in the Solar System, the deformation parameter $\epsilon$ in the Johannsen-Psaltis metric is to be thought as a free parameter to be determined by observations.

In the case of gravitational wave signals, we can just image that the line element in Eq.~(\ref{gmn}) is the vacuum static solution of some alternative theory of gravity. Since we want to remain as generic as possible, we do not want to specify (and actually we do not know) the exact field equations of the underlying theory. We can thus assume that the field equations are still given by Eq.~(\ref{eq-ee1}), but now $g^0_{\alpha\beta}$ is the Johannsen-Psaltis metric. We can then proceed as in the previous section and find the fundamental equations governing the axial perturbations generated by a test-particle moving along the geodesic of the background metric. Within this approach, axial perturbations are still described by Eq.~(\ref{eq-axial}),  but now the potential $V_l(r)$ and the source term $S_{lm}(t,r)$ are
\begin{widetext}
\be\label{eq-jp1}
V_l(r) &=& \left(1 - \frac{2M}{r}\right)
\left[\frac{l(l+1)}{r^2}\left(1 + \epsilon \frac{M^3}{r^3}\right) - \frac{6M}{r^3}\right] \, , \\
S_{lm}(t,r) &=& - 16\pi i \left(1 - \frac{2M}{r}\right)
\Bigg\{ r \frac{\partial}{\partial r}\left[\left(1 - \frac{2M}{r}\right) 
\left(1 + \epsilon\frac{M^3}{r^3}\right)^2 D_{lm}\right] 
- \left(1 - \frac{2M}{r}\right) 
\left(1 + \epsilon\frac{M^3}{r^3}\right)^2 Q_{lm}\Bigg\} \, .
\label{eq-jp1xx}
\ee
\end{widetext}
Of course, for $\epsilon=0$ we recover the standard case, namely Schwarzschild metric and general relativity. When $\epsilon \neq0$, we have a different wave form $Z_{lm}(t)$ and a different energy spectrum $dW/d\omega$. In this sense, we have introduced a phenomenological parametrization in the general relativity axial perturbations of a Schwarzschild BH and deviations from standard predictions are quantified in terms of $\epsilon$. However, such a deformation parameter $\epsilon$ is not the original $\epsilon$ in the Johannsen-Psaltis metric any more. The parameter appearing in Eq.~(\ref{eq-jp1}) and Eq.~(\ref{eq-jp1xx}) is a sort of effective parameter that takes deviations from the Einstein equations into account. The relation between this $\epsilon$ and the one appearing in the Johannsen-Psaltis metric and constrained from the electromagnetic spectrum of BH candidates~\cite{cfm1} is unknown at this level and therefore it is difficult to make any comparison between the two approaches. Of course, we could have also started from the very beginning from Eq.~(\ref{eq-axial}), writing some ad hoc correction in terms of a number of deformation parameters. However, this might have introduced additional arbitrariness and included unphysical deviations that cannot be obtained from any theory of gravity.

\begin{figure*}
\begin{center}
\includegraphics[type=pdf,ext=.pdf,read=.pdf,width=18.0cm]{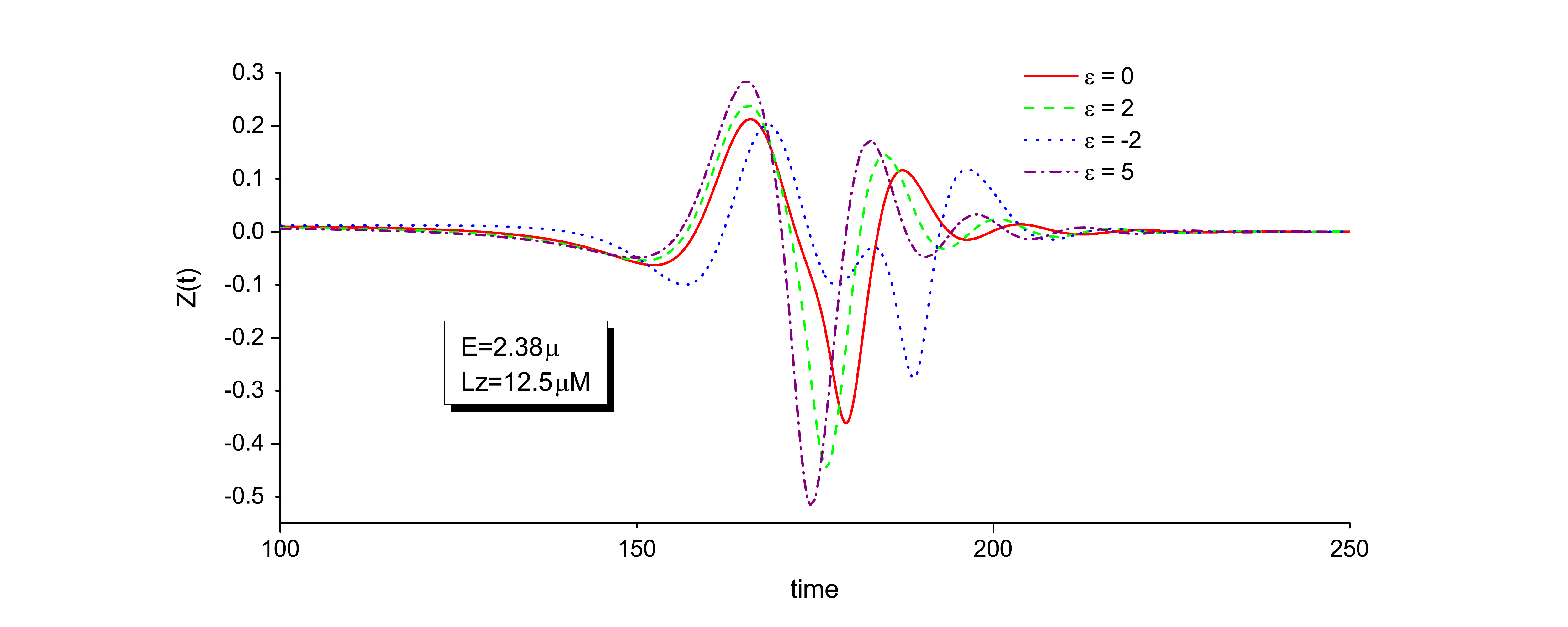}\\
\vspace{0.5cm}
\includegraphics[type=pdf,ext=.pdf,read=.pdf,width=18.0cm]{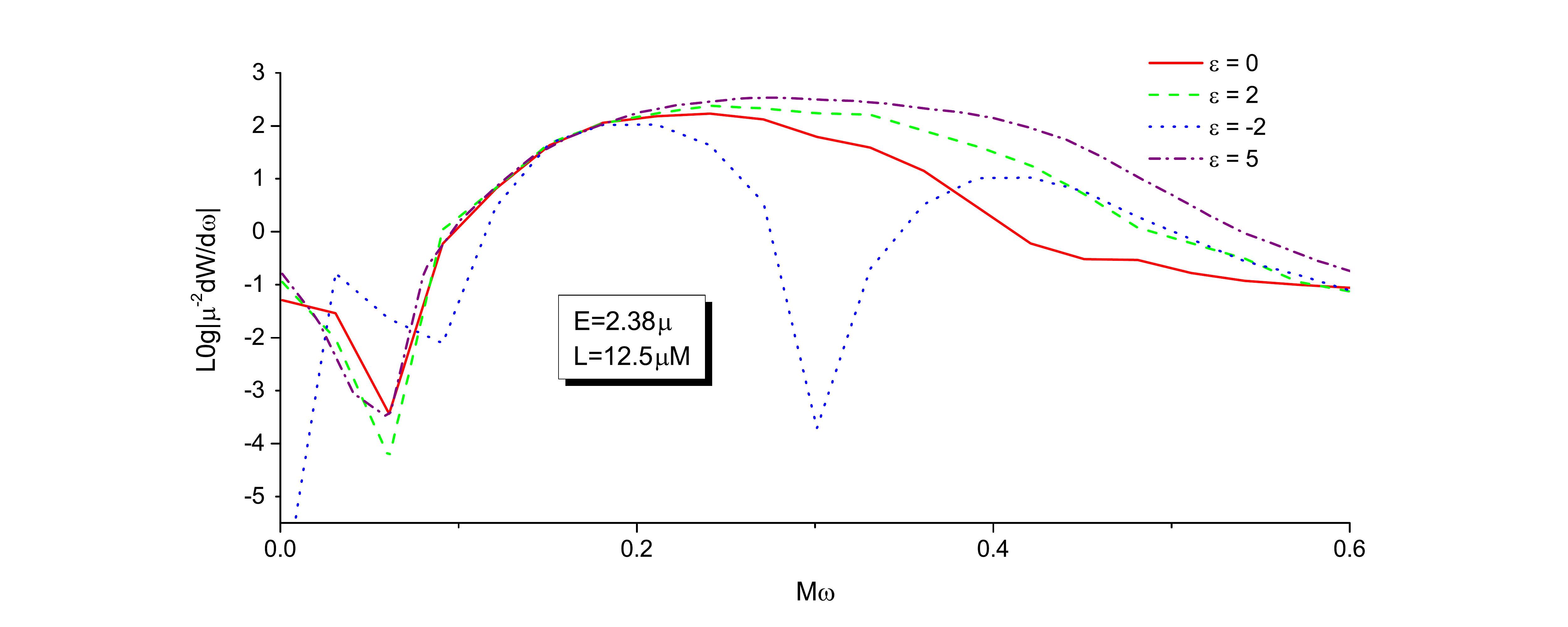}
\end{center}
\vspace{-0.5cm}
\caption{Wave form (top panel) and energy spectrum (bottom panel) produced by the excitation of axial quasi-normal modes of non-rotating BHs (Schwarzschild BH $\epsilon=0$, deformed BHs $\epsilon\neq0$) by a test particle with $E=2.38\mu$ and $L_z = 12.5\mu M$. See the text for more details.}
\label{fig1}
\end{figure*}

\begin{figure*}
\begin{center}
\includegraphics[type=pdf,ext=.pdf,read=.pdf,width=18.0cm]{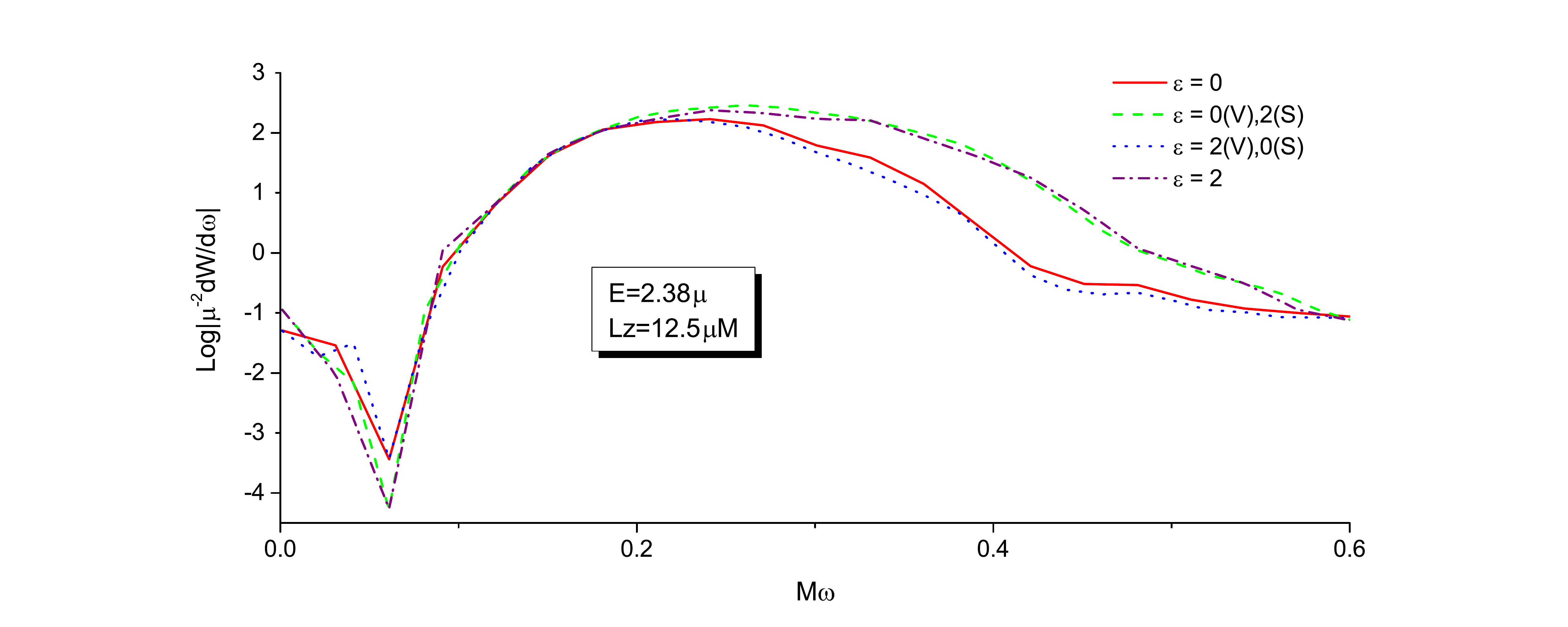}\\
\vspace{0.5cm}
\includegraphics[type=pdf,ext=.pdf,read=.pdf,width=18.0cm]{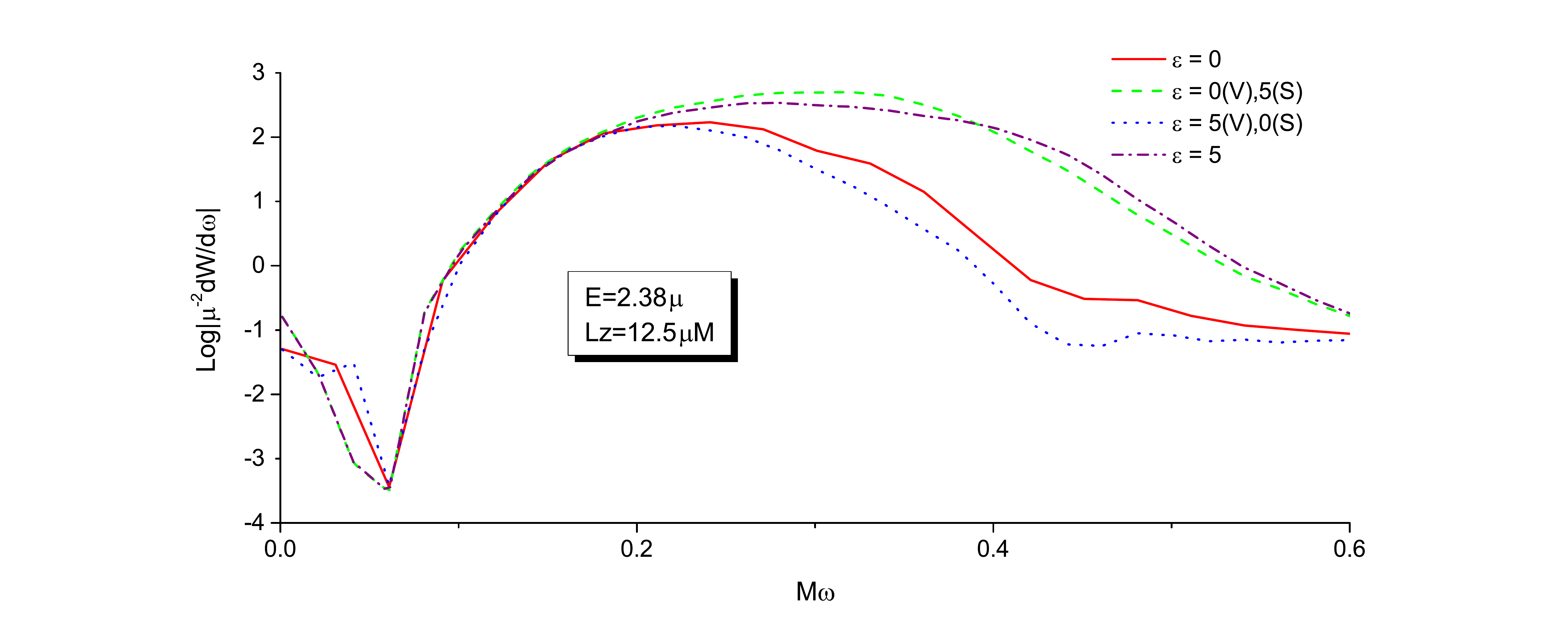}
\end{center}
\vspace{-0.5cm}
\caption{Energy spectrum in the case $\epsilon=0$ (red solid line), $\epsilon=0$ in $V_l(r)$ but $\epsilon\neq0$ in $S_{lm}(t,r)$ (green dashed line), $\epsilon\neq0$ in $V_l(r)$ but $\epsilon=0$ in $S_{lm}(t,r)$ (blue dotted line), and $\epsilon\neq0$ (violet dashed-dotted line). Here the test particle has $E=2.38\mu$ and $L_z = 12.5\mu M$. See the text for more details.}
\label{fig2}
\end{figure*}

\begin{figure*}
\begin{center}
\includegraphics[type=pdf,ext=.pdf,read=.pdf,width=18.0cm]{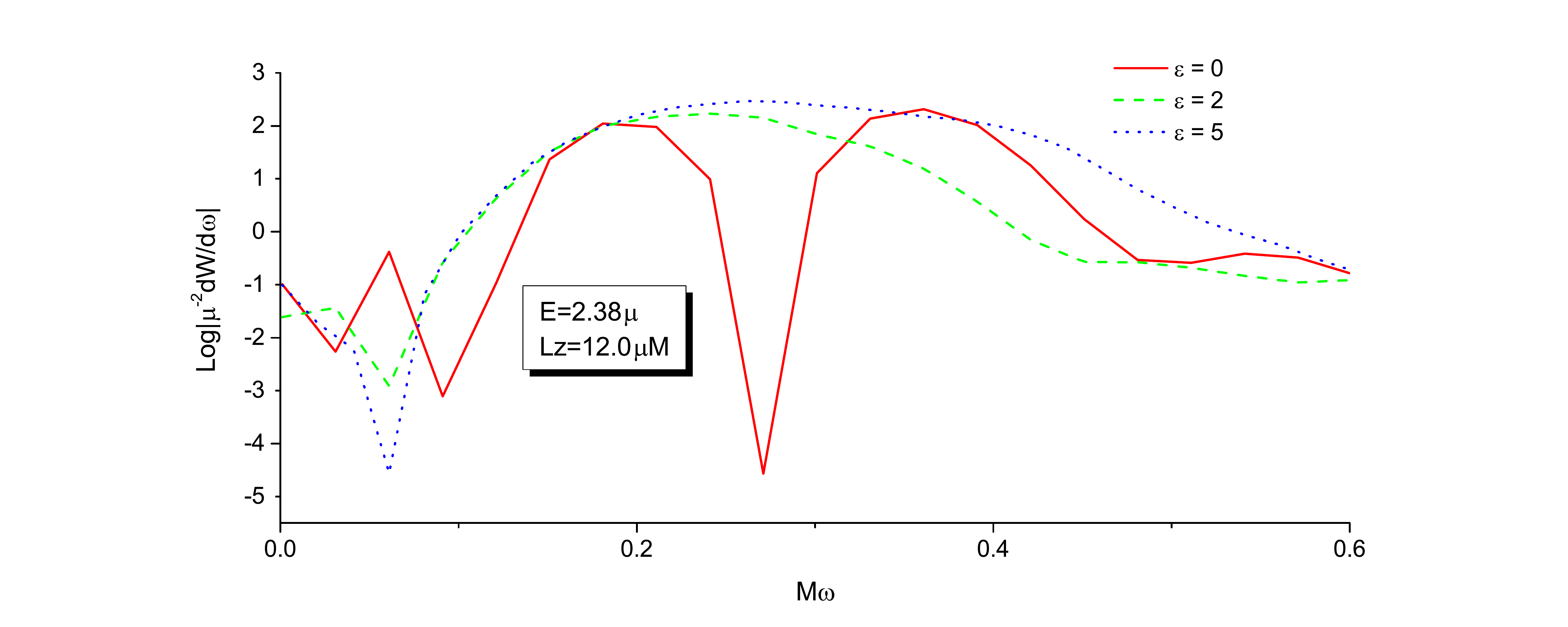}\\
\vspace{0.5cm}
\includegraphics[type=pdf,ext=.pdf,read=.pdf,width=18.0cm]{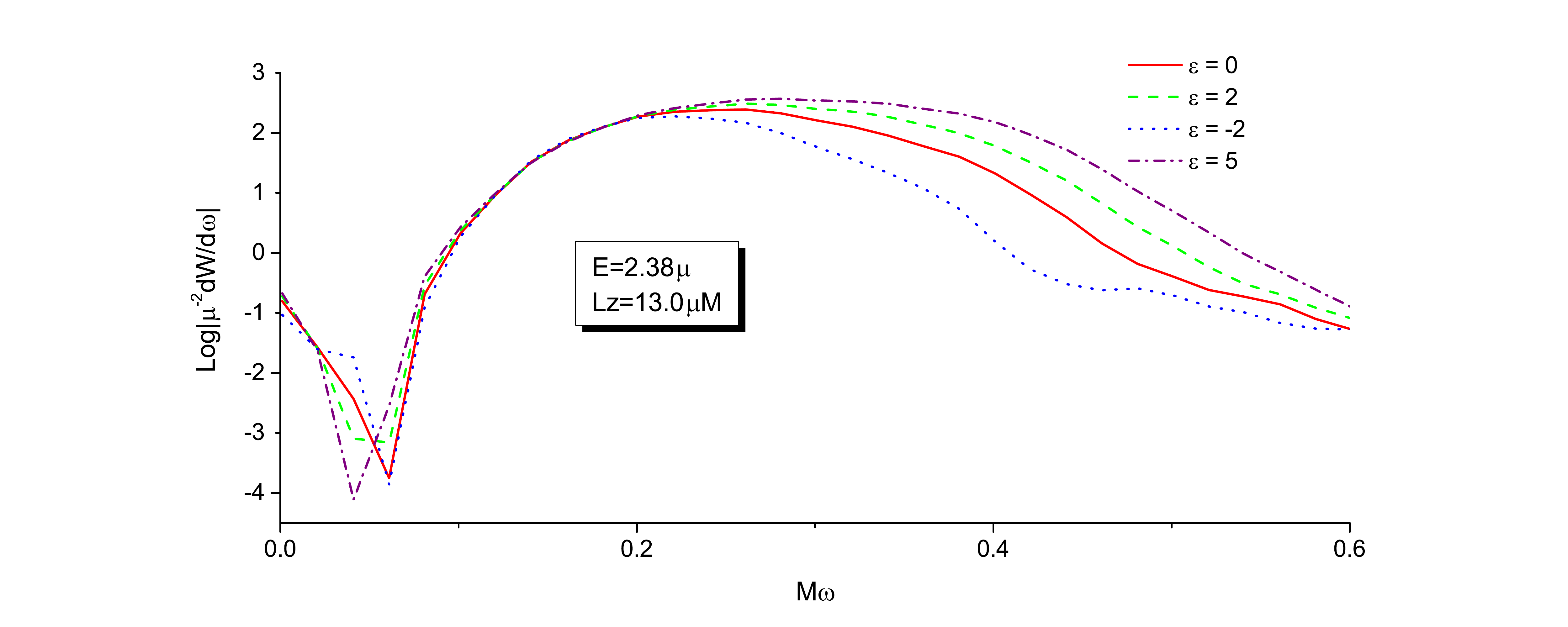}
\end{center}
\vspace{-0.5cm}
\caption{As in the bottom panel in Fig.~\ref{fig1}, but for a test particle with $E=2.38\mu$ and $L_z = 12.0\mu M$ (top panel) and with $E=2.38\mu$ and $L_z = 13.0\mu M$ (bottom panel). See the text for more details.}
\label{fig3}
\end{figure*}

\begin{figure*}
\begin{center}
\includegraphics[type=pdf,ext=.pdf,read=.pdf,width=7.5cm]{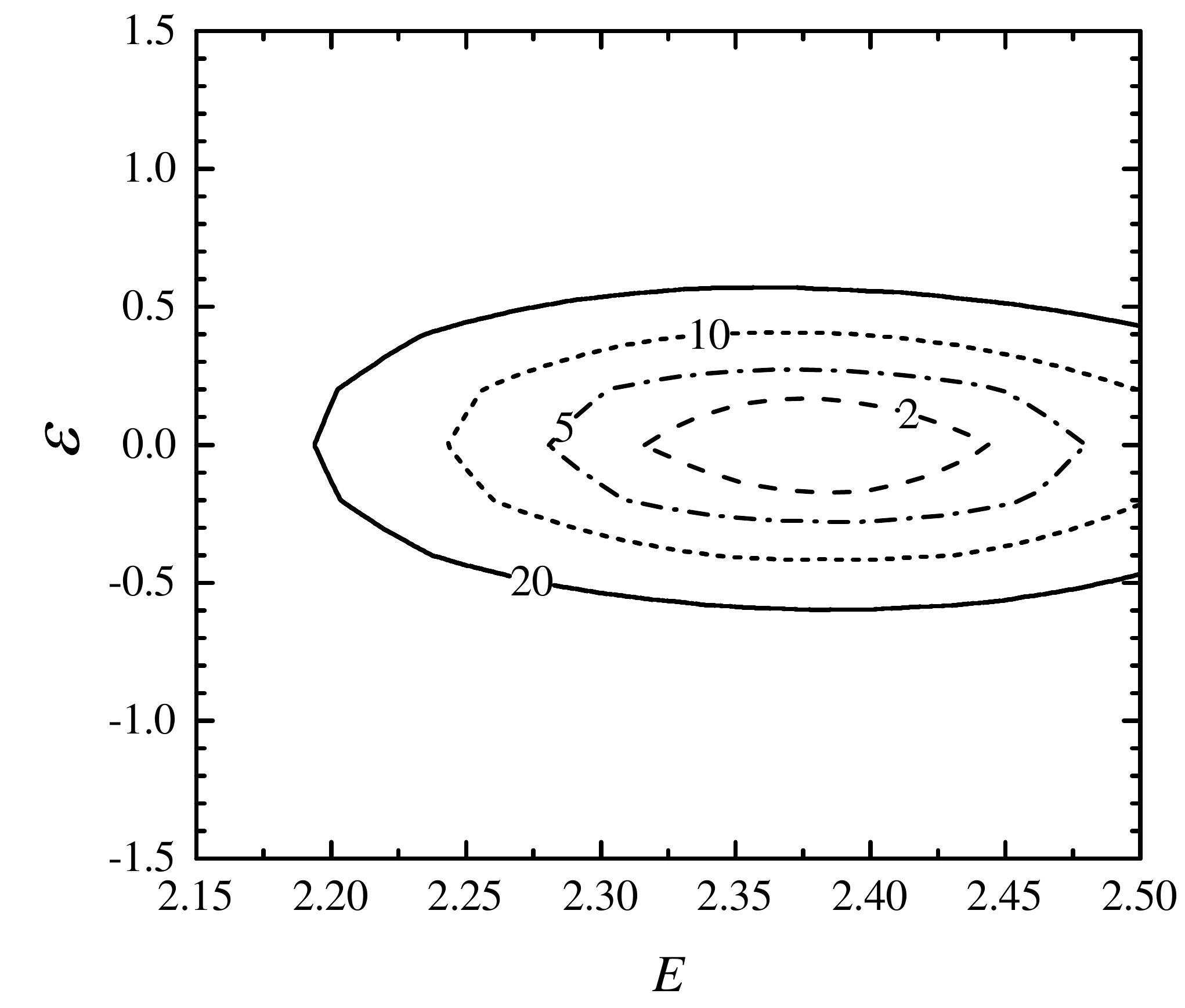}
\hspace{0.5cm}
\includegraphics[type=pdf,ext=.pdf,read=.pdf,width=7.5cm]{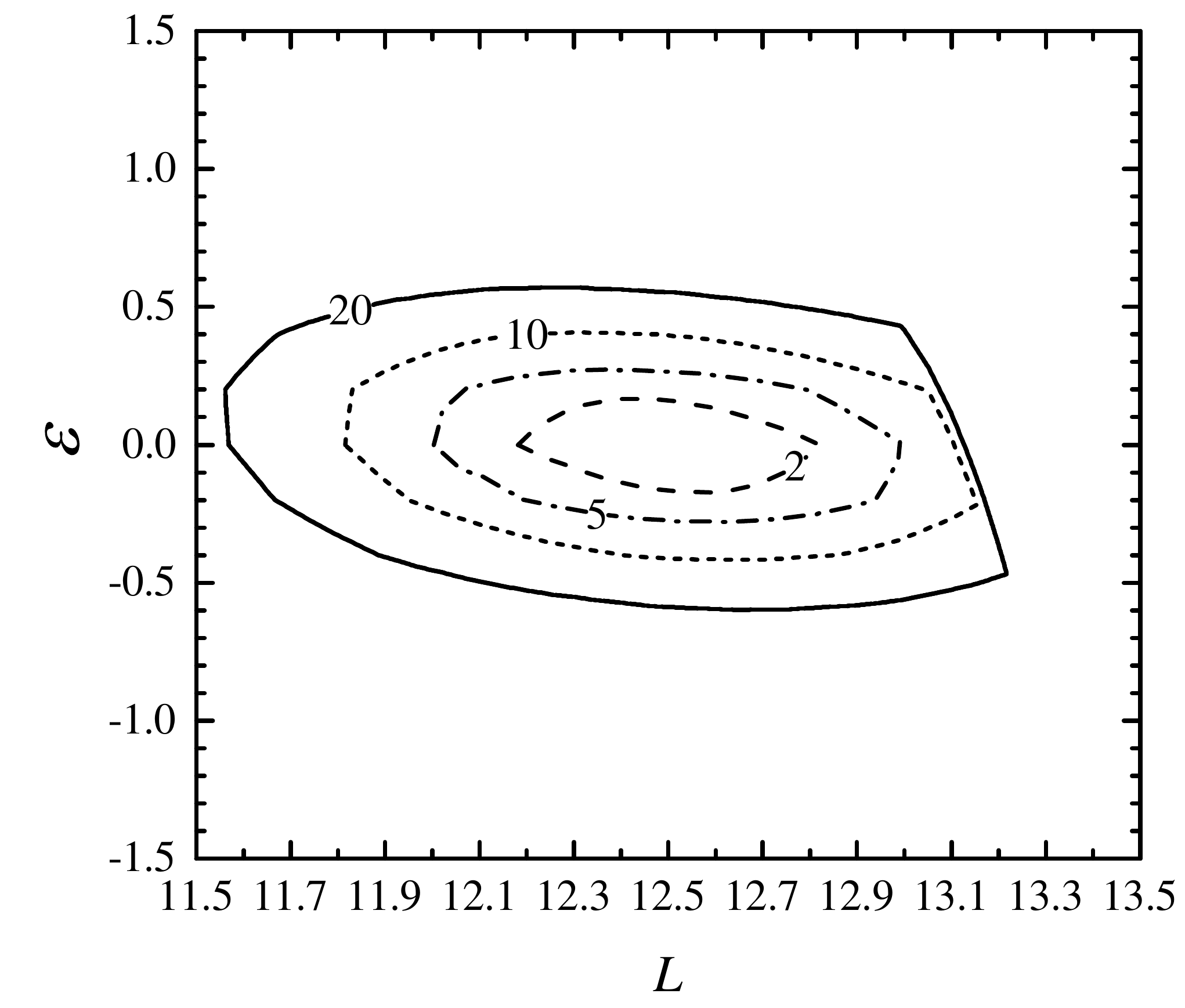}
\end{center}
\caption{Contour levels of $S$. The reference model has $\epsilon=0$ (Schwarzschild BH), $E = 2.38$, and $L_z = 12.5$ ($\mu = M = 1$). In the left panel, $S$ is minimized over $L_z$. In the right panel, $S$ is minimized over $E$. See the text for more details.}
\label{fig4}
\end{figure*}

\section{Simulations \label{s-sim}}

The numerical method to compute the wave form and the energy spectrum is the same as the one adopted in Ref.~\cite{kk1}. It consists of three steps.
\begin{enumerate}
\item We fix the parameters of the background metric ($\epsilon$) and of the particle ($E$ and $L_z$) and we compute the particle trajectory for a discrete number of time steps, ranging from a large $R$ to the turning point $R = R_t$. 
\item We compute the source term $S_{lm}(t,r)$ for each time and spatial point of the grid. Following Ref.~\cite{pullin}, we approximate the delta function by a narrow Gaussian
\be
\delta\left(r - R(t)\right) \approx \frac{\alpha}{\sqrt{\pi}} e^{-\alpha^2(r -R(t))^2} \, .
\ee
In our calculations, we use $\alpha = 5$, which is large enough not to affect our results.
\item We numerically solve Eq.~(\ref{eq-axial}). 
\end{enumerate}
With the above machinery, we compute the wave form at infinity, $Z_{lm}(t)$, and the energy spectrum, $dW/d\omega$, for test-particles moving in the gravitational field of Schwarzschild and deformed BHs. In our calculations, we only compute the signal of the multipole $l=2$ and $m=1$, which is the leading order term for axial perturbations.

Fig.~\ref{fig1} shows the wave form (top panel) and the energy spectrum (bottom panel) of the gravitational wave signal produced by a test-particle with $E=2.38\mu$ and $L_z=12.5\mu M$ moving in the spacetime of a Schwarzschild BH (red solid line) and of deformed non-rotating BHs with $\epsilon=\pm2$ and 5 (the case $\epsilon=-5$ has no turning point for this choice of $E$ and $L_z$). To quantify the relative contribution between the mode excitation and the quadrupole orbital emission, we have computed the energy spectrum with $\epsilon=0$ in $V_l(r)$ and $\epsilon\neq0$ in $S_{lm}(r)$ (quadrupole orbital emission) and then the case with $\epsilon\neq0$ in $V_l(r)$ and $\epsilon=0$ in $S_{lm}(r)$ (mode excitation), and compared with the Schwarzcshild and deformed BH spectra. The result is shown in Fig.~\ref{fig2}. It is evident that the gravitational wave signal is mainly determined by the geodesic motion of the test-particle in the background metric, while the contribution due to the mode excitation is smaller. This suggests that in the description of this kind of phenomena the field equations of the theory play a minor rule with respect to the geodesic motion. This may further justify our approach to plug the Johannsen-Psaltis metric in the Einstein equations: eventually the signal is mainly determined by the quadrupole moment formula and the partcle trajectory. The small contribution from the mode excitation can be easily understood if we notice that the maximum of the potential $V_l(r)$ is at $r_{max} \approx 3\;M$ and therefore $\epsilon (M/r_{max})^3 \ll 1$ for $\epsilon = O(1)$.

Fig.~\ref{fig3} show the energy spectrum in the case the test-particle has different $L_z$. In the top panel we set $L_z = 12.0\mu M$, in the bottom panel $L_z = 13.0\mu M$. We do not show the spectrum for $\epsilon=-2$ when $L_z = 12.0\mu M$ because in this case there is no turning point and the test-particle is swallowed by the BH.

In order to figure out whether an accurate measurement of the gravitational wave spectrum can distinguish different spacetimes and test general relativity, we can proceed as follows. We consider a reference model in which the spacetime is described by the Schwarzschild metric ($\epsilon=0$) and the test-particle has a certain specific energy and axial component of the specific angular momentum. Such a reference model can be compared to another model, in which the metric has a deformation parameter $\epsilon$, and the specific energy and axial component of the angular momentum are, respectively, $E$ and $L_z$, by evaluating the following function (for the sake of simplicity, we set $M=\mu=1$)
\be\label{eq-sss}
S(\epsilon,E,L_z) = \sum_i 
\left[\frac{\log \left(\frac{dW}{d\omega}\right)_i
\left(\epsilon,E,L_z\right) - \log \left(\frac{dW}{d\omega}\right)_i^{\rm ref}}{C \log 
\left(\frac{dW}{d\omega}\right)_i^{\rm ref}}\right]^2 \, , \nonumber\\
\ee
where $\left(dW/d\omega\right)_i^{\rm ref}$ is the energy spectrum of the reference model at the frequency $\omega_i$, $\left(dW/d\omega\right)_i \left(\epsilon,E,L_z\right)$ is the energy spectrum of the model with parameters $\left(\epsilon,E,L_z\right)$, and $C$ is a constant that we have quite arbitrarily set to 0.3. Eq.~(\ref{eq-sss}) clearly looks like a $\chi^2$ and $C$ as the error. We do not call $S$ $\chi^2$ simply because we do not want to perform a rigorous analysis, which would require a more detailed discussion beyond the scope of our explorative work.

Fig.~\ref{fig4} shows the contour level of $S$ assuming that the test-particle in the reference model has $E=2.38\mu$ and $L_z=12.5M\mu$. The sum is performed over 41 frequencies, from $\omega_0 = 0$ to $\omega_{40} = 0.6/M$. In the left panel, $S$ is minimized over $L_z$. In the right panel, $S$ is minimized over $E$. If we identified $S$ with $\chi^2$, $S = 3.5$, 8.0, and 14.2 would correspond to the probability interval designated as 1-, 2-, and 3-standard deviations for three degrees of freedom. From this contour levels, it seems that the measurement of the energy spectrum can test the background metric and constrain $\epsilon$. There is no correlation between the deformation parameter $\epsilon$ and the parameters of the test-particle $E$ and $L_z$. This is a good result if we want to test BH candidates, but we have to consider that we are restricting our attention to non-rotating BHs only. The typical problem to test BH candidates is the strong correlation between the spin and possible deviations from the Kerr solution. We leave the extension to rotating BHs to a future work.

\section{Summary and conclusions \label{s-c}}

Astrophysical BH candidates are supposed to be the Kerr BH of general relativity, but any observational confirmation is still lacking and the same Einstein's theory has not yet been tested in strong gravitational fields. The Kerr paradigm can be potentially verified by studying the electromagnetic and gravitational signals emitted from these systems. The radiation emitted by the gas in the inner region of the accretion disk is affected by relativistic phenomena and the study of the electromagnetic spectrum of a BH candidate can thus provide information on the background metric close to the compact object. Gravitational waves produced by perturbations in the spacetime geometry around BH candidates are determined by both the background metric and the field equations of the gravity theory.

In this paper, we have considered the axial perturbations generated by a test-particle moving in the gravitational field of non-rotating BHs. Starting from the Johannsen-Psaltis parametrization, we have derived some effective equations governing the evolutions of the gravitational wave signal and we have compute the associated energy spectrum. Deviations from the Schwarzschild predictions in general relativity are quantified by an effective deformation parameter $\epsilon$. Because of the dependence of the gravitational wave signal on the field equations of the underlying theory, it is not possible to directly compare the results from electromagnetic and gravitational spectrum. In the former case, experiments can really constrain the deformation parameter appearing in the Johannsen-Psaltis metric. With the gravitational wave approach, the final parameter has also absorbed possible deviations from the Einstein equations.

We find that the gravitational wave signal is mainly determined by the geodesic motion of the particle, while deviations in the axial quasi-normal modes due to a different background are smaller. Employing a simple analysis with the $S$ function in Eq.~(\ref{eq-sss}), we have compared different models to figure out if and how the energy spectrum can constrain the deformation parameter $\epsilon$ and thus test general relativity. Our result is promising: as shown in Fig.~\ref{fig4}, it seems that the measurement of the gravitational wave spectrum can distinguish different spacetimes, constraining $\epsilon$ and determining the parameters of the test-particle $E$ and $L_z$. However, our discussion is limited to non-rotating BHs. The typical problem to test BH candidates is the strong correlation between the estimate of the spin and of the deformation parameters. We leave the analysis of rotating BHs to a future work, which is fundamental to understand if and how this approach can test BH candidates.

%%%%%%%%%%%%%%%%%%%%%%%%%%%%%%%

\begin{acknowledgments}
We thanks Kostas Kokkotas for useful discussions and suggestions. This work was supported by the NSFC grant No.~11305038, the Shanghai Municipal Education Commission grant No.~14ZZ001, the Thousand Young Talents Program, and Fudan University. C.B. acknowledges also support from the Alexander von Humboldt Foundation.
\end{acknowledgments}

%%%%%%%%%%%%%%%%%%%%%%%%%%%%%%


\begin{thebibliography}{99}

\bibitem{will} 
  C.~M.~Will,
  %``The Confrontation between general relativity and experiment,''
  Living Rev.\ Rel.\  {\bf 9}, 3 (2006)
  [gr-qc/0510072].

\bibitem{cosmo} 
  B.~Jain and J.~Khoury,
  %``Cosmological Tests of Gravity,''
  Annals Phys.\  {\bf 325}, 1479 (2010)
  [arXiv:1004.3294 [astro-ph.CO]].

\bibitem{bh1} 
  C.~Bambi,
  %``Testing the Kerr black hole hypothesis,''
  Mod.\ Phys.\ Lett.\ A {\bf 26}, 2453 (2011)
  [arXiv:1109.4256 [gr-qc]].  
  
\bibitem{bh2} 
  C.~Bambi,
  %``Testing the space-time geometry around black hole candidates with the available radio and X-ray data,''
  Astron.\ Rev.\  {\bf 8}, 4 (2013)
  [arXiv:1301.0361 [gr-qc]].

\bibitem{new1}
  G.~Dvali and C.~Gomez,
  %``Black Hole's Quantum N-Portrait,''
  Fortsch.\ Phys.\  {\bf 61}, 742 (2013)
  [arXiv:1112.3359 [hep-th]].  
  
\bibitem{new2}
  G.~Dvali and C.~Gomez,
  %``Black Hole's 1/N Hair,''
  Phys.\ Lett.\ B {\bf 719}, 419 (2013)
  [arXiv:1203.6575 [hep-th]].  

\bibitem{new3} 
  S.~B.~Giddings,
  %``Possible observational windows for quantum effects from black holes,''
  Phys.\ Rev.\ D {\bf 90}, 124033 (2014)
  [arXiv:1406.7001 [hep-th]].

\bibitem{hair1}
  B.~Carter,
  %``Axisymmetric Black Hole Has Only Two Degrees of Freedom,''
  Phys.\ Rev.\ Lett.\  {\bf 26}, 331 (1971).
  
\bibitem{hair2}
  D.~C.~Robinson,
  %``Uniqueness of the Kerr black hole,''
  Phys.\ Rev.\ Lett.\  {\bf 34}, 905 (1975).

\bibitem{luca} 
  L.~Baiotti, I.~Hawke, P.~J.~Montero, F.~Loffler, L.~Rezzolla, N.~Stergioulas, J.~A.~Font and E.~Seidel,
  %``Three-dimensional relativistic simulations of rotating neutron star collapse to a Kerr black hole,''
  Phys.\ Rev.\ D {\bf 71}, 024035 (2005)
  [gr-qc/0403029].

\bibitem{collapse}
  R.~H.~Price,
  %``Nonspherical perturbations of relativistic gravitational collapse. 1. Scalar and gravitational perturbations,''
  Phys.\ Rev.\ D {\bf 5}, 2419 (1972).

\bibitem{charge}
  C.~Bambi, A.~D.~Dolgov and A.~A.~Petrov,
  %``Black holes as antimatter factories,''
  JCAP {\bf 0909}, 013 (2009)
  [arXiv:0806.3440 [astro-ph]].
  
\bibitem{disk}
  C.~Bambi, D.~Malafarina and N.~Tsukamoto,
  %``Note on the effect of a massive accretion disk in the measurements of black hole spins,''
  Phys.\ Rev.\ D {\bf 89}, 127302 (2014)
  [arXiv:1406.2181 [gr-qc]].

\bibitem{rr}
  C.~E.~Rhoades and R.~Ruffini,
  %``Maximum mass of a neutron star,''
  Phys.\ Rev.\ Lett.\  {\bf 32}, 324 (1974).

\bibitem{maoz}
  E.~Maoz,
  %``Dynamical constraints on alternatives to massive black holes in galactic nuclei,''
  Astrophys.\ J.\  {\bf 494}, L181 (1998)
  [astro-ph/9710309].

\bibitem{eh1}
  R.~Narayan and J.~E.~McClintock,
  %``Advection-Dominated Accretion and the Black Hole Event Horizon,''
  New Astron.\ Rev.\  {\bf 51}, 733 (2008)
  [arXiv:0803.0322 [astro-ph]].
  
\bibitem{eh2}
  M.~A.~Abramowicz, W.~Kluzniak and J.~-P.~Lasota,
  %``No observational proof of the black hole event-horizon,''
  Astron.\ Astrophys.\  {\bf 396}, L31 (2002)
  [astro-ph/0207270].
  
\bibitem{eh3}  
  C.~Bambi,
  %``A note on the observational evidence for the existence of event horizons in astrophysical black hole candidates,''
  The Scientific World Journal {\bf 2013}, 204315 (2013)
  [arXiv:1205.4640 [gr-qc]].

\bibitem{cb1} 
  C.~Bambi and E.~Barausse,
  %``Constraining the quadrupole moment of stellar-mass black-hole candidates with the continuum fitting method,''
  Astrophys.\ J.\  {\bf 731}, 121 (2011)
  [arXiv:1012.2007 [gr-qc]]. 
  
\bibitem{cb2} 
  C.~Bambi,
  %``Constraint on the quadrupole moment of super-massive black hole candidates from the estimate of the mean radiative efficiency of AGN,''
  Phys.\ Rev.\ D {\bf 83}, 103003 (2011)
  [arXiv:1102.0616 [gr-qc]].  
  
\bibitem{cb3} 
  C.~Bambi,
  %``A code to compute the emission of thin accretion disks in non-Kerr space-times and test the nature of black hole candidates,''
  Astrophys.\ J.\  {\bf 761}, 174 (2012)
  [arXiv:1210.5679 [gr-qc]].
  
\bibitem{cb4} 
  C.~Bambi,
  %``Testing the space-time geometry around black hole candidates with the analysis of the broad K$\alpha$ iron line,''
  Phys.\ Rev.\ D {\bf 87}, 023007 (2013)
  [arXiv:1211.2513 [gr-qc]]. 
  
\bibitem{cb5} 
  C.~Bambi,
  %``Probing the space-time geometry around black hole candidates with the resonance models for high-frequency QPOs and comparison with the continuum-fitting method,''
  JCAP {\bf 1209}, 014 (2012)
  [arXiv:1205.6348 [gr-qc]].  
  
\bibitem{b055} 
  C.~Bambi,
  %``Measuring the Kerr spin parameter of a non-Kerr compact object with the continuum-fitting and the iron line methods,''
  JCAP {\bf 1308}, 055 (2013)
  [arXiv:1305.5409 [gr-qc]].  
  
\bibitem{jp1} 
  T.~Johannsen and D.~Psaltis,
  %``Testing the No-Hair Theorem with Observations in the Electromagnetic Spectrum. III. Quasi-Periodic Variability,''
  Astrophys.\ J.\  {\bf 726}, 11 (2011)
  [arXiv:1010.1000 [astro-ph.HE]].  
  
\bibitem{jp2} 
  T.~Johannsen and D.~Psaltis,
  %``Testing the No-Hair Theorem with Observations in the Electromagnetic Spectrum. IV. Relativistically Broadened Iron Lines,''
  Astrophys.\ J.\  {\bf 773}, 57 (2013)
  [arXiv:1202.6069 [astro-ph.HE]].   
  
\bibitem{corr1} 
  C.~Bambi,
  %``Testing the Kerr-nature of stellar-mass black hole candidates by combining the continuum-fitting method and the power estimate of transient ballistic jets,''
  Phys.\ Rev.\ D {\bf 85}, 043002 (2012)
  [arXiv:1201.1638 [gr-qc]].  
  
\bibitem{corr2} 
  C.~Bambi,
  %``Attempt to find a correlation between the spin of stellar-mass black hole candidates and the power of steady jets: relaxing the Kerr black hole hypothesis,''
  Phys.\ Rev.\ D {\bf 86}, 123013 (2012)
  [arXiv:1204.6395 [gr-qc]]. 
  
\bibitem{corr3} 
  J.~Jiang, C.~Bambi and J.~F.~Steiner,
  %``Using iron line reverberation and spectroscopy to distinguish Kerr and non-Kerr black holes,''
  JCAP {\bf 1505}, 025 (2015)
  [arXiv:1406.5677 [gr-qc]].   
  
\bibitem{corr4} 
  J.~Jiang, C.~Bambi and J.~F.~Steiner,
  %``Testing the Kerr Nature of Black Hole Candidates using Iron Line Spectra in the CPR Framework,''
  arXiv:1504.01970 [gr-qc].  
  
\bibitem{e1} 
  P.~S.~Joshi, D.~Malafarina and R.~Narayan,
  %``Distinguishing black holes from naked singularities through their accretion disc properties,''
  Class.\ Quant.\ Grav.\  {\bf 31}, 015002 (2014)
  [arXiv:1304.7331 [gr-qc]].
  
\bibitem{e2} 
  C.~Bambi and D.~Malafarina,
  %``K$\alpha$ iron line profile from accretion disks around regular and singular exotic compact objects,''
  Phys.\ Rev.\ D {\bf 88}, 064022 (2013)
  [arXiv:1307.2106 [gr-qc]]. 

\bibitem{e3} 
  C.~Bambi,
  %``Broad K? iron line from accretion disks around traversable wormholes,''
  Phys.\ Rev.\ D {\bf 87}, 084039 (2013)
  [arXiv:1303.0624 [gr-qc]]. 

\bibitem{cfm1} 
  L.~Kong, Z.~Li and C.~Bambi,
  %``Constraints on the spacetime geometry around 10 stellar-mass black hole candidates from the disk's thermal spectrum,''
  Astrophys.\ J.\  {\bf 797}, 78 (2014)
  [arXiv:1405.1508 [gr-qc]].  

\bibitem{cfm2}
  C.~Bambi,
  %``Note on the Cardoso-Pani-Rico parametrization to test the Kerr black hole hypothesis,''
  Phys.\ Rev.\ D {\bf 90}, 047503 (2014)
  [arXiv:1408.0690 [gr-qc]].
 
\bibitem{f1} 
  N.~Tsukamoto, Z.~Li and C.~Bambi,
  %``Constraining the spin and the deformation parameters from the black hole shadow,''
  JCAP {\bf 1406}, 043 (2014)
  [arXiv:1403.0371 [gr-qc]].   
   
\bibitem{f2} 
  C.~Bambi,
  %``Constraining the Cardoso-Pani-Rico metric with future observations of SgrA$^*$,''
  Class.\ Quant.\ Grav.\  {\bf 32}, 065005 (2015)
  [arXiv:1409.0310 [gr-qc]].    
   
\bibitem{gw1}
  K.~Glampedakis and S.~Babak,
  %``Mapping spacetimes with LISA: inspiral of a test-body in a `quasi-Kerr'
  %field,''
  Class.\ Quant.\ Grav.\  {\bf 23}, 4167 (2006)
  [arXiv:gr-qc/0510057].

\bibitem{gw2}
  L.~Barack and C.~Cutler,
  %``Using LISA EMRI sources to test off-Kerr deviations in the geometry of
  %massive black holes,''
  Phys.\ Rev.\  D {\bf 75}, 042003 (2007)
  [arXiv:gr-qc/0612029].

\bibitem{gw3}
  T.~A.~Apostolatos, G.~Lukes-Gerakopoulos and G.~Contopoulos,
  %``How to Observe a Non-Kerr Spacetime Using Gravitational Waves,''
  Phys.\ Rev.\ Lett.\  {\bf 103}, 111101 (2009)
  [arXiv:0906.0093 [gr-qc]].
  
\bibitem{psaltis} 
  D.~Psaltis, D.~Perrodin, K.~R.~Dienes and I.~Mocioiu,
  %``Kerr Black Holes are Not Unique to General Relativity,''
  Phys.\ Rev.\ Lett.\  {\bf 100}, 091101 (2008)
  [Phys.\ Rev.\ Lett.\  {\bf 100}, 119902 (2008)]
  [arXiv:0710.4564 [astro-ph]].    

\bibitem{gwp1} 
  S.~Gossan, J.~Veitch and B.~S.~Sathyaprakash,
  %``Bayesian model selection for testing the no-hair theorem with black hole ringdowns,''
  Phys.\ Rev.\ D {\bf 85}, 124056 (2012)
  [arXiv:1111.5819 [gr-qc]].

\bibitem{gwp2} 
  N.~Yunes and F.~Pretorius,
  %``Fundamental Theoretical Bias in Gravitational Wave Astrophysics and the Parameterized Post-Einsteinian Framework,''
  Phys.\ Rev.\ D {\bf 80}, 122003 (2009)
  [arXiv:0909.3328 [gr-qc]]. 
  
\bibitem{kk1} 
  V.~Ferrari and K.~D.~Kokkotas,
  %``Scattering of particles by neutron stars: Time evolutions for axial perturbations,''
  Phys.\ Rev.\ D {\bf 62}, 107504 (2000)
  [gr-qc/0008057].   
  
\bibitem{zerilli} 
  F.~J.~Zerilli,
  %``Effective potential for even parity Regge-Wheeler gravitational perturbation equations,''
  Phys.\ Rev.\ Lett.\  {\bf 24}, 737 (1970).  

\bibitem{jpm} 
  T.~Johannsen and D.~Psaltis,
  %``A Metric for Rapidly Spinning Black Holes Suitable for Strong-Field Tests of the No-Hair Theorem,''
  Phys.\ Rev.\ D {\bf 83}, 124015 (2011)
  [arXiv:1105.3191 [gr-qc]].
  
\bibitem{pullin} 
  J.~Ruoff, P.~Laguna and J.~Pullin,
  %``Excitation of neutron star oscillations by an orbiting particle,''
  Phys.\ Rev.\ D {\bf 63}, 064019 (2001)
  [gr-qc/0005002].  

\end{thebibliography}
\end{document}